\documentclass[twocolumn]{aastex62}

\usepackage{graphicx}
\usepackage{amssymb}
\usepackage{amsmath}
\usepackage{natbib}
\usepackage{wrapfig}
\usepackage[utf8]{inputenc}
\usepackage[T1]{fontenc}
\newcommand{\masyr}{\>{\rm mas}\,{\rm yr}^{-1}}

\newcommand{\muw}{\mu_{W}}
\newcommand{\mun}{\mu_{N}}

\newcommand{\hst}{{\it HST}}
\newcommand{\gaia}{{\it Gaia}}

\submitjournal{ApJ}
\accepted{February 8th, 2019}

\shorttitle{Magellanic Bridge Proper Motions}
\shortauthors{Zivick et al.}

\begin{document}

\title{The Proper Motion Field Along the Magellanic Bridge: a New Probe of the LMC-SMC Interaction}


\correspondingauthor{Paul Zivick}
\email{pjz2cf@virginia.edu}

\author[0000-0001-9409-3911]{Paul Zivick}
\affiliation{Department of Astronomy, University of Virginia, 530 McCormick Road, Charlottesville, VA 22904, USA}

\author{Nitya Kallivayalil}
\affiliation{Department of Astronomy, University of Virginia, 530 McCormick Road, Charlottesville, VA 22904, USA}

\author{Gurtina Besla}
\affiliation{Steward Observatory, University of Arizona, 933 North Cherry Avenue, Tucson, AZ 85721, USA}

\author{Sangmo Tony Sohn}
\affiliation{Space Telescope Science Institute, 3700 San Martin Drive, Baltimore, MD 21218, USA}

\author{Roeland P. van der Marel}
\affiliation{Space Telescope Science Institute, 3700 San Martin Drive, Baltimore, MD 21218, USA}
\affiliation{Center for Astrophysical Sciences, Department of Physics \& Astronomy, Johns Hopkins University, Baltimore, MD 21218, USA}

\author{Andr\'es del Pino}
\affiliation{Space Telescope Science Institute, 3700 San Martin Drive, Baltimore, MD 21218, USA}

\author{Sean T. Linden}
\affiliation{Department of Astronomy, University of Virginia, 530 McCormick Road, Charlottesville, VA 22904, USA}

\author{Tobias K. Fritz}
\affiliation{Instituto de Astrofisica de Canarias, Calle Via Lactea s/n, E-38205 La Laguna, Tenerife, Spain}
\affiliation{Universidad de La Laguna, Dpto. Astrofisica, E-38206 La Laguna, Tenerife, Spain}

\author{J. Anderson}
\affiliation{Space Telescope Science Institute, 3700 San Martin Drive, Baltimore, MD 21218, USA}

\begin{abstract}
We present the first detailed kinematic analysis of the proper motions (PMs) of stars in the Magellanic Bridge, from both the \gaia\ Data Release 2 catalog and from \textit{Hubble Space Telescope} Advanced Camera for Surveys data. For the \gaia\ data, we identify and select two populations of stars in the Bridge region, young main sequence (MS) and red giant stars. The spatial locations of the stars are compared against the known H {\small I} gas structure, finding a correlation between the MS stars and the H {\small I} gas. In the \hst\ fields our signal comes mainly from an older MS and turn-off population, and the proper motion baselines range between $\sim 4$ and 13 years. The PMs of these different populations are found to be consistent with each other, as well as across the two telescopes. When the absolute motion of the Small Magellanic Cloud is subtracted out, the residual Bridge motions display a general pattern of pointing away from the Small Magellanic Cloud towards the Large Magellanic Cloud. We compare in detail the kinematics of the stellar samples against numerical simulations of the interactions between the Small and Large Magellanic Clouds, and find general agreement between the kinematics of the observed populations and a simulation in which the Clouds have undergone a recent direct collision.

\end{abstract}

\section{Introduction}

Stretched between the Small and Large Magellanic Clouds (SMC, LMC respectively) lies the Magellanic Bridge, originally identified as an overdensity of H {\small I} gas by \cite{hindman63}. Given the proximity of the two dwarfs, tidal interactions between them were a clear potential explanation, and in time, models of the Magellanic system demonstrated this generally accepted paradigm \citep[e.g.,][]{besla12,diaz12}. Measurements of the relative motions of the SMC and LMC suggest that their most recent interaction likely occurred $\sim$ 150 Myr ago, with an impact parameter of $< 10$ kpc \citep{zivick18}. This implies that the Magellanic Bridge was formed via both hydrodynamic and tidal interactions \citep{besla12}.

One additional prediction of the models is the presence of both in situ star formation as well as older, tidally stripped stars. Even before the formal predictions, a population of young stars associated with the Bridge was observed by \cite{irwin85}, with a follow-up study by \cite{demers98}, that would be consistent with in situ star formation. \cite{harris07} further examined this young population, hoping to use the star formation history to constrain the interactions between the Clouds. The existence of young stellar objects in the region \citep[e.g.,][]{sewilo13} and the strong correlation between young stars and the H {\small I} overdensities \citep[e.g.,][]{skowron14} helped confirm the in situ formation scenario. 

Only recently has there been evidence for the presence of older SMC stars in the Bridge. Using a combination of the WISE and 2MASS surveys, \cite{bagheri13} identified red giant branch (RGB) stars scattered around the Bridge region (later confirmed by \cite{noel13}). Spectroscopic follow up of targets in the region by \cite{carrera17} found the stars to be older than 1 Gyr and with metallicities consistent with having formed in the outer regions of the SMC. The stripping of the SMC was also observed by \cite{belokurov17} in \gaia\ DR1 data where they found two spatially distinct structures, separated by multiple degrees, made up of young main sequence stars and RR Lyrae stars. 

These structures and their kinematic properties play an important role in understanding the interaction history between the Clouds. Different factors governing this interaction history have been explored in the literature, including varying the masses of the dwarfs, the impact parameters of the interaction, duration of the interaction time, and other factors  \citep[e.g.,][]{besla12,diaz12}, each one providing a set of predictions. Understanding the 3D structure of the Bridge can help to constrain these formation scenarios \citep[e.g.,][]{belokurov17}, and detailed kinematic information will aid in further improving those constraints. Recent efforts have found a trend of stars moving from the SMC to the LMC in the plane of the sky \citep[e.g.,][]{schmidt18} and outward motions on the eastern edge of the SMC distinct from the dwarf's motion \citep{oey18}, supporting the idea that material has been stripped from the SMC, but no detailed kinematic analysis of the Bridge has yet been published.

In this paper, we present the first detailed analysis of the proper-motion (PM) field of stars in the Magellanic Bridge, and directly compare these PMs to predictions from simulations of the interaction history of the Clouds. 
We use the recently published \gaia\ Data Release 2 catalog (\citealt{gaia16a}; \citealt{gaiadr218b}) in combination with \hst\ data to examine the kinematic structure in the Bridge. We examine both young and old stellar populations 
in the Bridge region. We treat each population separately and consider for the young stars the H {\small I} gas structure for potential correlations. For the comparisons with theory, we use the models presented in \cite{besla12}.

The paper is organized as follows. In Section \ref{sec:data} we discuss the selection criteria applied to the \gaia\ data as well as the analysis and calculation of the PMs from the \hst\ data. This data is transformed into a model-ready comparison frame, described at the beginning of Section \ref{sec:analysis}. From there we examine the spatial and kinematic differences between the young and old populations and the young stellar population's spatial correlation with the H {\small I} gas. We close Section \ref{sec:analysis} by making direct comparisons with simulations of the past interactions of the Clouds. Finally, in Section \ref{sec:dnc} we summarize our findings and their implications for our understanding of the Magellanic system.

\section{Data Selection}\label{sec:data}

\subsection{\gaia\ DR2 Data}\label{ssec:gaiadata}

\begin{figure}
\begin{center}
 \includegraphics[width=3.4in]{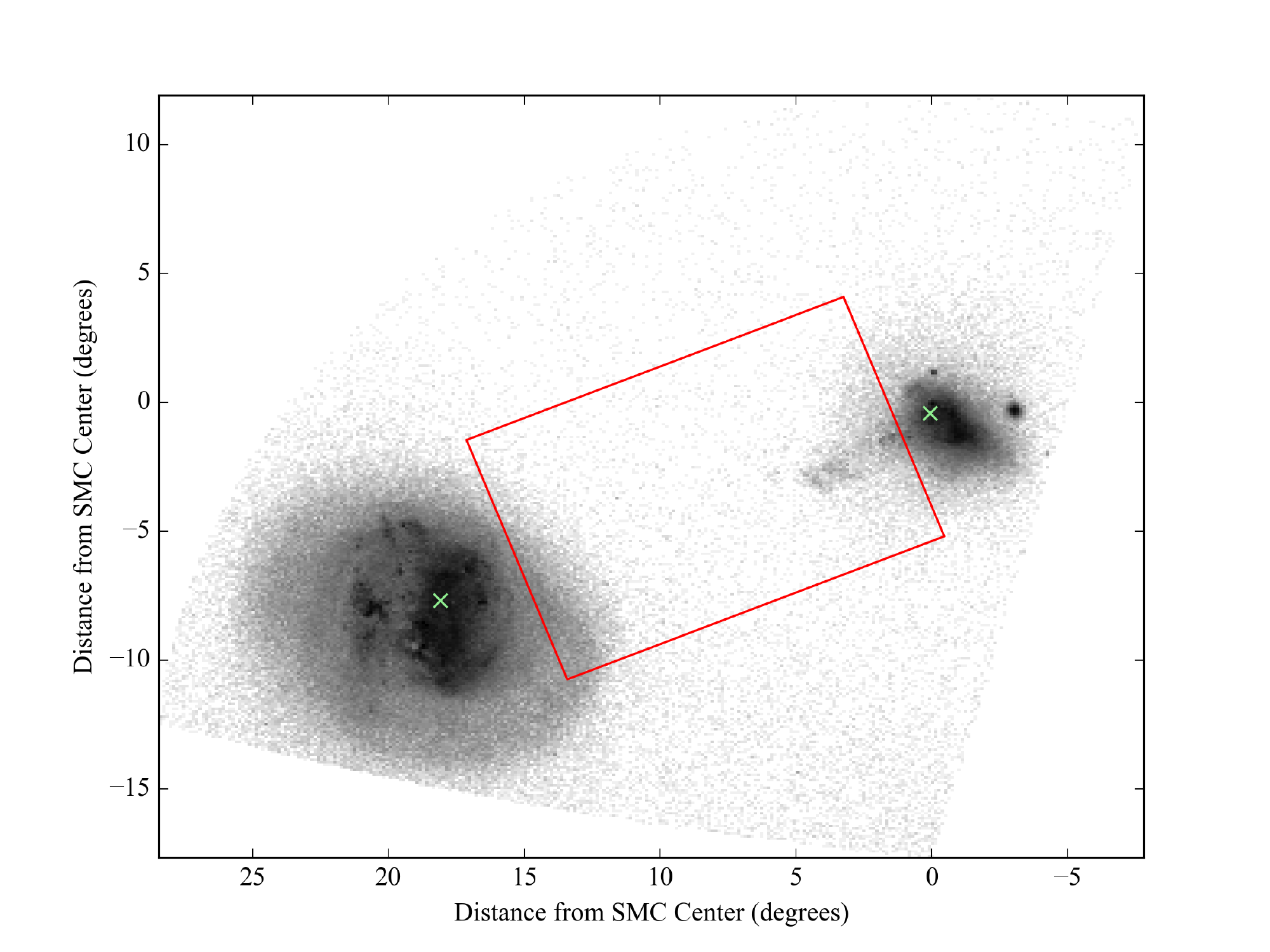} 
 \caption{\gaia\ source density count around the Magellanic System with cuts made as described in \ref{ssec:gaiadata} for astrometric quality. The green crosses mark the locations of the assumed centers of the LMC and SMC, and the red box indicates the area examined further for Bridge dynamics. (0,0) is defined as the kinematic center of the SMC.}
   \label{fig:mc}
\end{center}
\end{figure}

From the \gaia\ database, we select all stars within the vicinity of the Clouds (the exact area is shown in Figure \ref{fig:mc}) using \texttt{pygacs}\footnote{\small \url{https://github.com/Johannes-Sahlmann/pygacs}}. We begin with a simple parallax cut of $\omega < 0.2$ mas in order to remove foreground MW stars. Next we apply the following cut to the renormalized unit weight error (RUWE) as described in the \gaia\ technical note GAIA-C3-TN-LU-LL-124-01:
\begin{equation}
    \frac{\sqrt{\chi^2 / (N-5)}} {u_{0}(G,C)} < 1.40,
\end{equation}
where $\chi^2$ is the \gaia\ property \texttt{astrometric\_chi2\_al}, $N$ is \texttt{astrometric\_n\_good\_obs\_al}, and $u_{0}$ is an empirically-derived normalization factor that uses \texttt{phot\_g\_mean\_mag} ($G$) and \texttt{bp\_rp} ($C$). We additionally apply a cut for the color excess of the stars, as described in \cite{gaiadr2hrd} by Equation C.2. As we are concerned with the better astrometrically behaved stars, primarily the bright stars, and to provide another check to avoid MW contamination, we select stars brighter than $G < 17$, leading to the final source densities in Figure \ref{fig:mc}.

\begin{figure}
\begin{center}
 \includegraphics[width=3.4in]{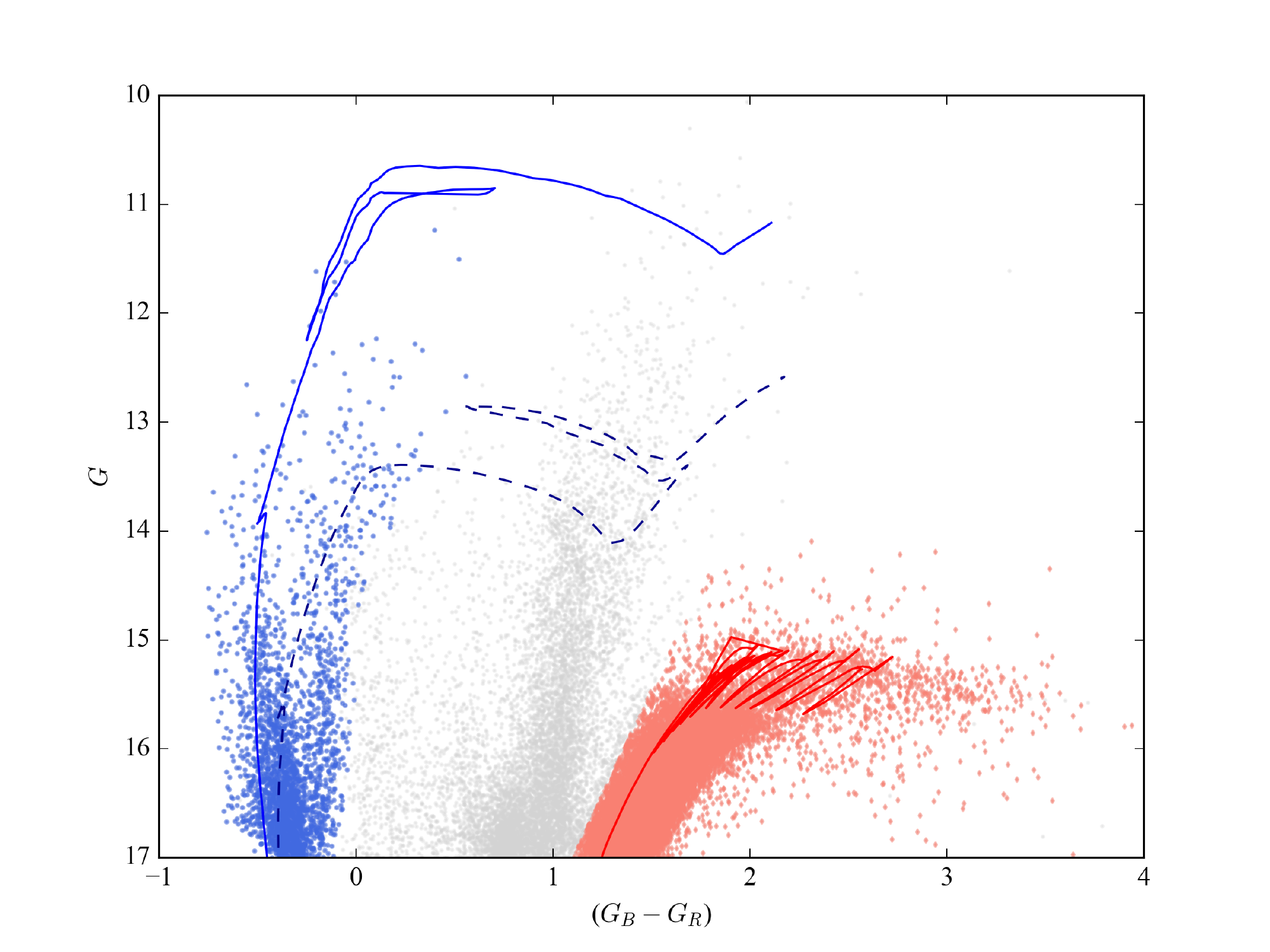} 
 \caption{Color-Magnitude Diagram of the selected Bridge region. All stars in the region are marked in gray. The blue colored points indicate the stars selected by our mask as main sequence stars, and the orange-red colored points indicate the red giant mask. From left to right the \texttt{PARSEC} isochrones are 10 Myr (solid blue line), 30 Myr (dashed dark blue line), and 800 Myr (solid red line), all more metal-rich than [M/H] $> -0.65$.}
   \label{fig:cmdsel}
\end{center}
\end{figure}

\begin{figure}
\begin{center}
 \includegraphics[width=3.4in]{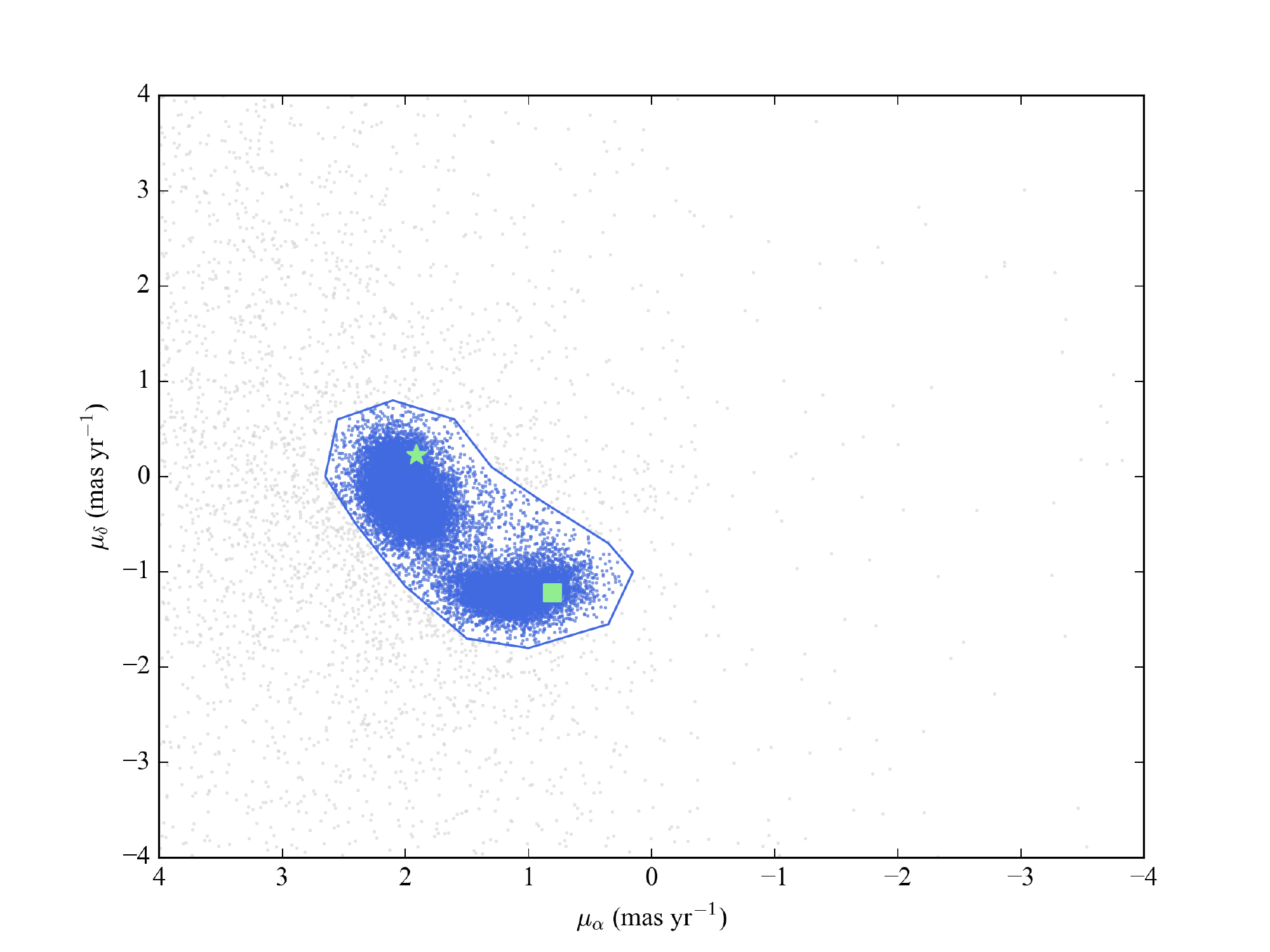} 
 \caption{Proper motion diagram for the selected Bridge stars. All stars present in the region are marked in gray. The blue lines indicate the PM region identified as belonging to the Magellanic system. The light green square indicates the PM of the SMC and the light green star indicates the PM of the LMC.}
 
   \label{fig:pmsel}
\end{center}
\end{figure}

From this initial catalog we select a smaller area for closer examination, stretching from the eastern edge of the SMC to the western edge of the LMC. These boundaries are marked in red in Figure \ref{fig:mc}. From this region, we apply two more criteria in the location of the stars in the color-magnitude diagram (CMD) and their PMs. For the CMD, first we de-redden our sample of \gaia\ stars, using \cite{gaiadr218b} and \cite{schlegel98}. Using the de-reddened CMD, we select main sequence (MS) and red giant (RG) stars as indicated in Figure \ref{fig:cmdsel}. We provide for reference three \texttt{PARSEC} isochrones \citep{marigo17}, the two in blue at 10 and 30 Myr and the one in red at 800 Myr. We note that we are not attempting a rigorous fit to the stellar populations, but instead we use these to highlight the likely populations belonging to the Clouds. The two young isochrones do appear to trace distinct MS populations, especially above $G < 15$. An examination of the spatial and kinematic properties of the two populations revealed no apparent difference, so for the comparison to both the older population and numerical modeling, all MS stars will be categorized together. 
For the PM selection (see Figure \ref{fig:pmsel}), we select all stars in and around the two dense regions, with each region belonging to one of the Clouds, with the systemic motions marked in light green. With this cut in PM, we allow for stars originating from the LMC to be included in the sample. Given the large overdensity in Figure \ref{fig:pmsel}, it is likely that many of the stars, especially those spatially overlapping with the LMC, are of LMC-origins. However, due to the uncertainty in assigning a definite membership to any given star, we keep this broader PM selection to provide as much relevant information regarding the Bridge as possible. Our final sample only includes stars that pass both of these cuts.

\begin{figure}
\begin{center}
 \includegraphics[width=3.4in]{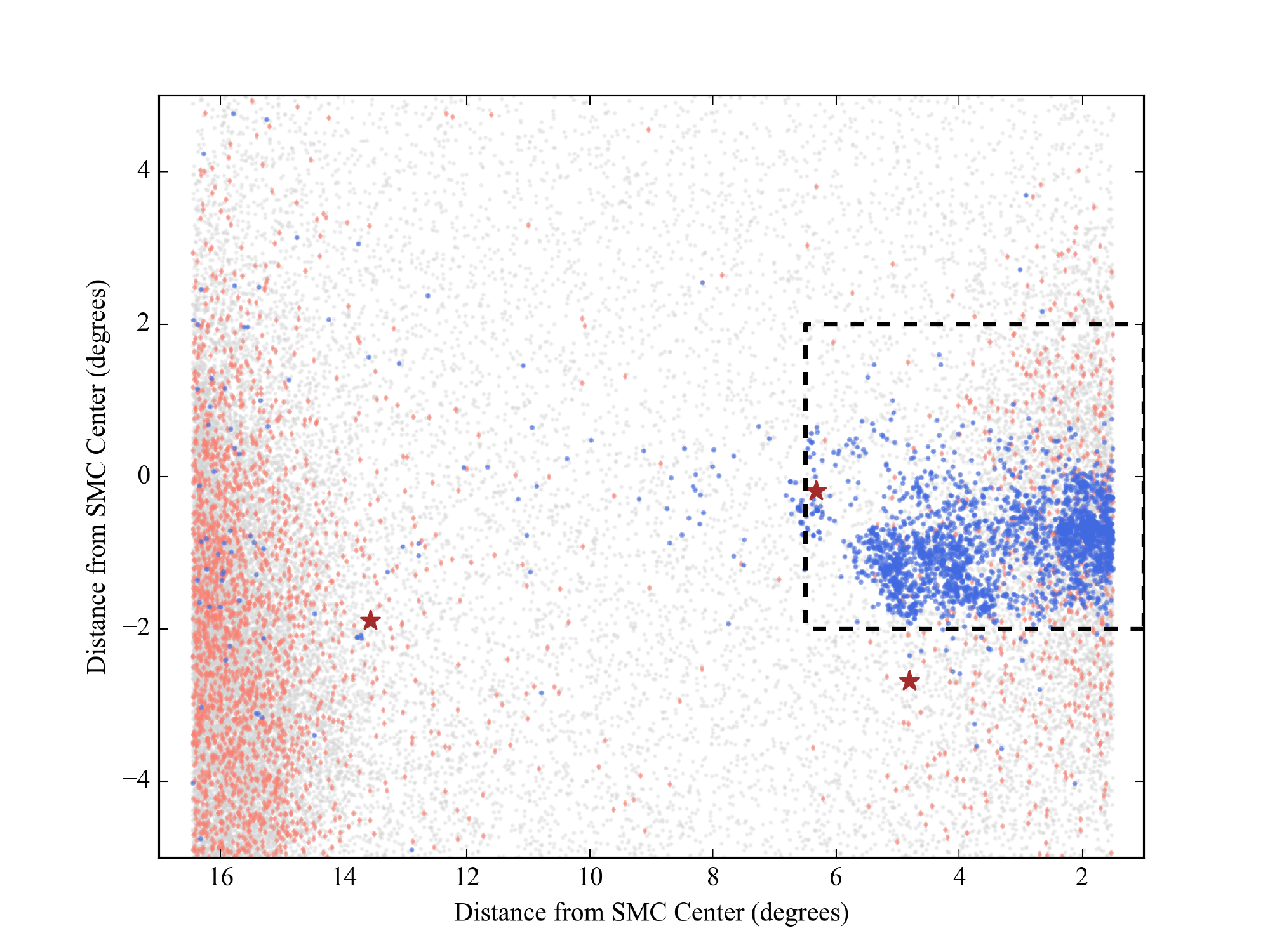} 
 \caption{Bridge region with the main sequence (MS) population marked in blue and the red giant population (RG) marked in orange-red. The frame has been rotated such that the $x$-axis now lies along the line between the assumed centers for the LMC and SMC, where (0,0) is the center of the SMC. The area comparable to Model 2 (discussed in \ref{ssec:comp}) has been outlined in black for easier comparison, and the locations of the \hst\ fields are marked with brown stars. The RG population has been randomly subsampled down to the level of the MS stars to allow for easier comparison of the spatial correlations.}
   \label{fig:modelpos}
\end{center}
\end{figure}

A subsample of roughly 3,000 MS stars and 20,000 RG stars pass our astrometric and CMD-based cuts. Examining the physical location of the stars in this sample, we see that the selected MS stars trace the expected Bridge structure while the RG stars primarily trace the broader SMC and LMC structure, although some RG stars are scattered throughout the Bridge area (Figure \ref{fig:modelpos}). For easier viewing, the RG population has been randomly subsampled to the same number as MS stars. 
Our selected area does include part of the region identified as possessing LMC substructure in the RG population in \gaia\ DR2 data by \cite{belokurov19}. This substructure, roughly located in the bottom left of our Figure \ref{fig:modelpos}, can be slightly seen, but we ascribe most of the difference to our brighter magnitude cut of $G < 17$ removing much of the signal in addition to the subsampling done for display purposes.
From here, we begin to examine the kinematic properties of the stars as they relate to the larger Magellanic system.

\subsection{\hst\ Data}\label{ssec:hstdata}

%
\begin{deluxetable*}{lcccccccc}
\tablecolumns{11}
\tablewidth{0pc}
\tablecaption{\hst\ Target Fields and Observations
              \label{t:bgfields}}
\tablehead{
\colhead{}              & \colhead{R.A.}    & \colhead{Decl.}   & \multicolumn{3}{c}{Epoch 1}                                                     & \colhead{} & \multicolumn{2}{c}{Epoch 2 (Prog. ID 13834)} \\
\cline{4-6} \cline{8-9} \\
\colhead{Target Fields} & \colhead{(J2000)} & \colhead{(J2000)} & \colhead{Prog. ID} & \colhead{Epoch} & \colhead{Exp. Time (s)\tablenotemark{a}} & \colhead{} & \colhead{Epoch} & \colhead{Exp. Time (s)\tablenotemark{a}} 
}
\startdata
\hst-BG1 & 02:04:11.2 & $-$76:16:11.5 &    12286 & 2011.49 & 2132 & & 2015.43 & 9126 \\
\hst-BG2 & 02:30:41.6 & $-$73:53:43.3 & \phn9488 & 2003.20 & 2400 & & 2015.21 & 8757 \\
\hst-BG3 & 04:21:05.0 & $-$74:02:26.9 & \phn9488 & 2002.72 & 1800 & & 2015.68 & 9246 \\
\enddata
\tablenotetext{a}{Total exposure time of the F775W observations used for astrometric analysis.}
\end{deluxetable*}
%

In addition to the \gaia\ PMs, we measured PMs of stars in the Magellanic Bridge using \hst\ data. 
We searched the \hst\ archive for existing 
deep imaging located along the Magellanic Bridge and found three fields. The  
characteristics of these fields are summarized in Table~\ref{t:bgfields}. 
The first-epoch data for the three fields were obtained for \hst\ programs 
to study the cosmic shear or Lyman-break galaxies at high redshift.
The second-epoch data were obtained through our \hst\ program GO-13834 
(PI: van der Marel) to measure PMs. We used the same observational setup 
(i.e. telescope pointing, orientation, detector, and filters) as the 
first-epoch observations. For the astrometric analysis we used the F775W filter data, and to construct 
CMDs of our target fields that may help in 
identifying stars along the Magellanic bridge against Galactic foreground 
contamination, we obtained F606W exposures during our second-epoch 
observations. 

We measured the PMs of stars in our target fields using the same 
technique as used in \citet{sohn15,sohn16}. Readers interested in the 
details of the PM measurement process are referred to those papers.
In short, we created high-resolution stacked images by combining our 
second-epoch data, identified stars and background galaxies from these 
stacks, constructed templates for stars and galaxies, determined 
template-based positions of stars and galaxies on images in each epoch, 
and measured displacements in positions of stars with respect to the 
background galaxies between the two epochs. We also measured photometry 
for each star in our target fields in the F606W and F775W bands.  
To do this, we used \textit{AstroDrizzle} \citep{gonzaga12} to combine 
images for each field per filter and measured the flux within 
aperture radius of 0.1 mas (i.e., 4 ACS/WFC pixels) from the center of 
each stars. Aperture corrections were carried out to infinity following 
the method by \citet{sirianni05}. The photometry was then calibrated  
to the ACS/WFC VEGAMAG system using the time-dependent zero points 
provided by the STScI webpage. 

%
%
\begin{figure*}
\gridline{ \fig{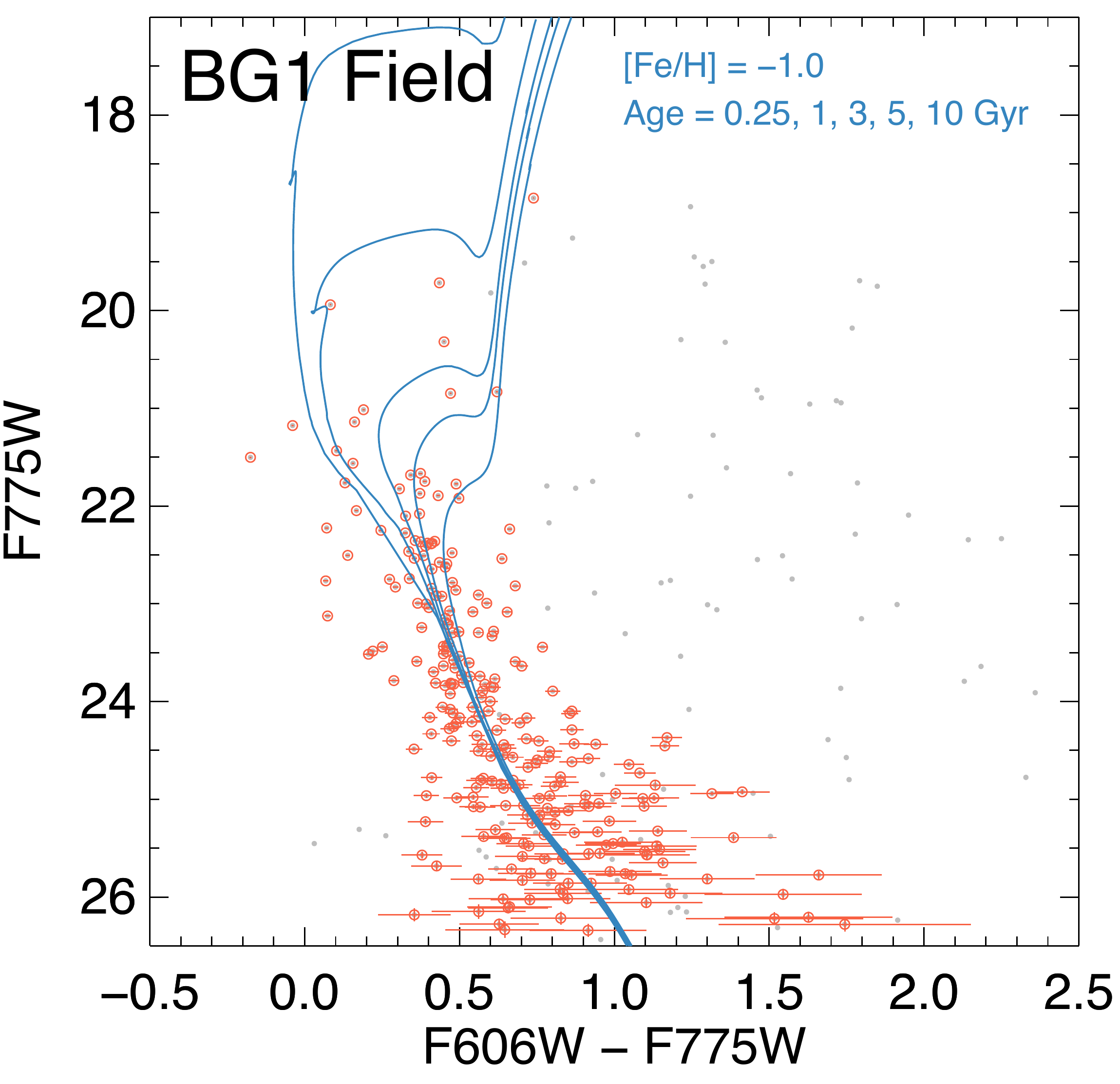}{0.33\textwidth}{} 
           \fig{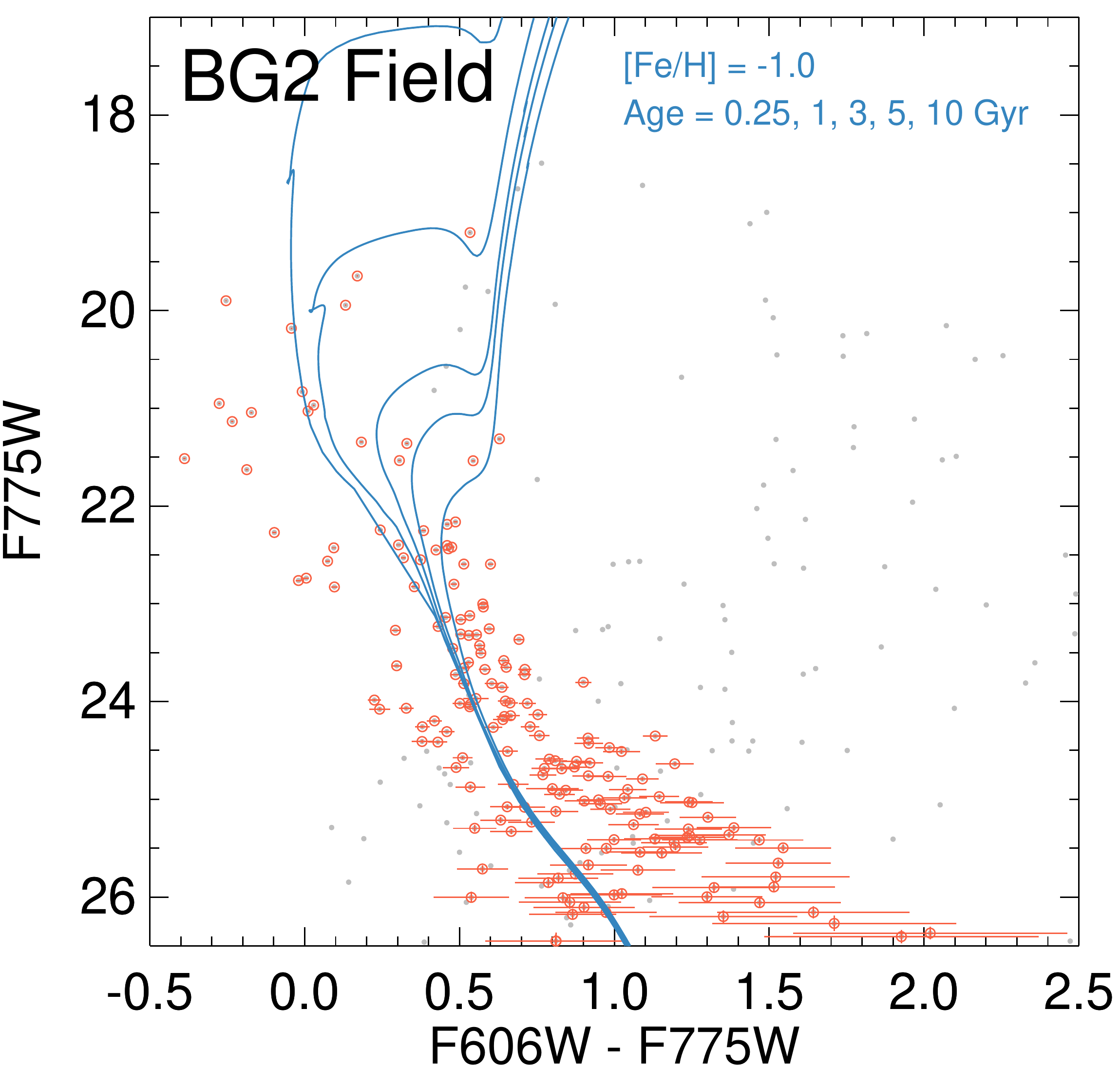}{0.33\textwidth}{}
           \fig{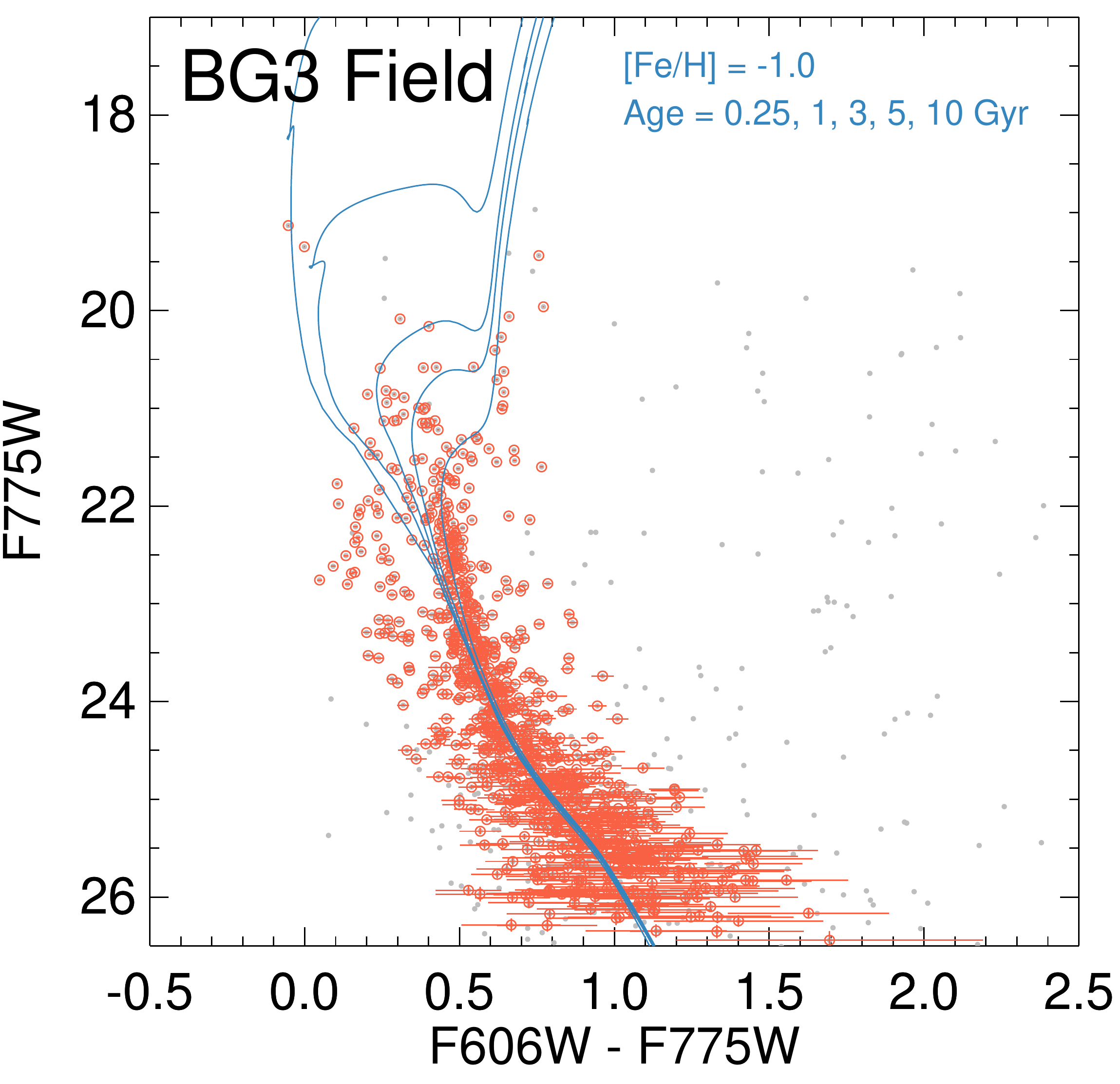}{0.33\textwidth}{}
            }
\gridline{ \fig{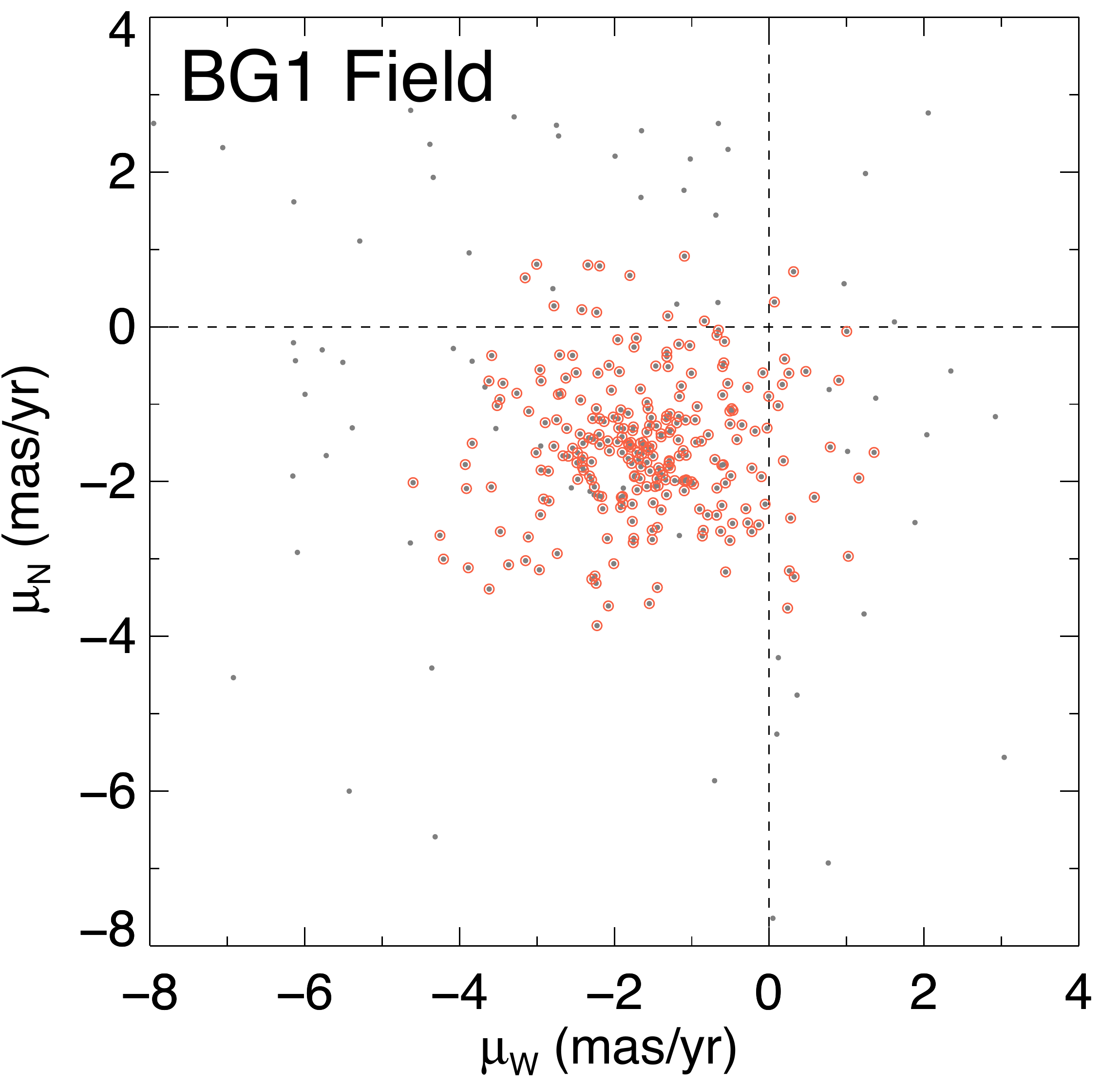}{0.33\textwidth}{} 
           \fig{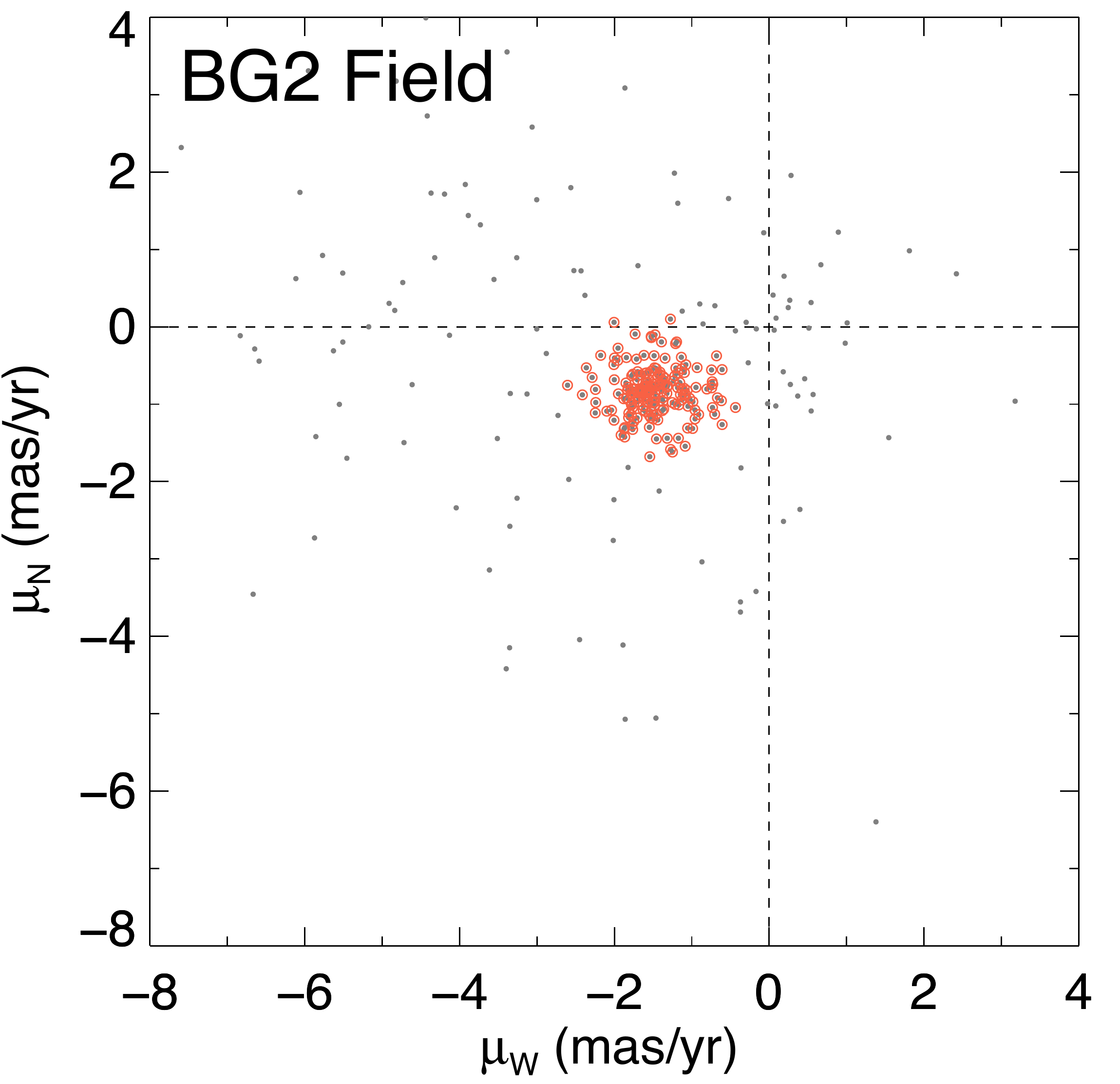}{0.33\textwidth}{}
           \fig{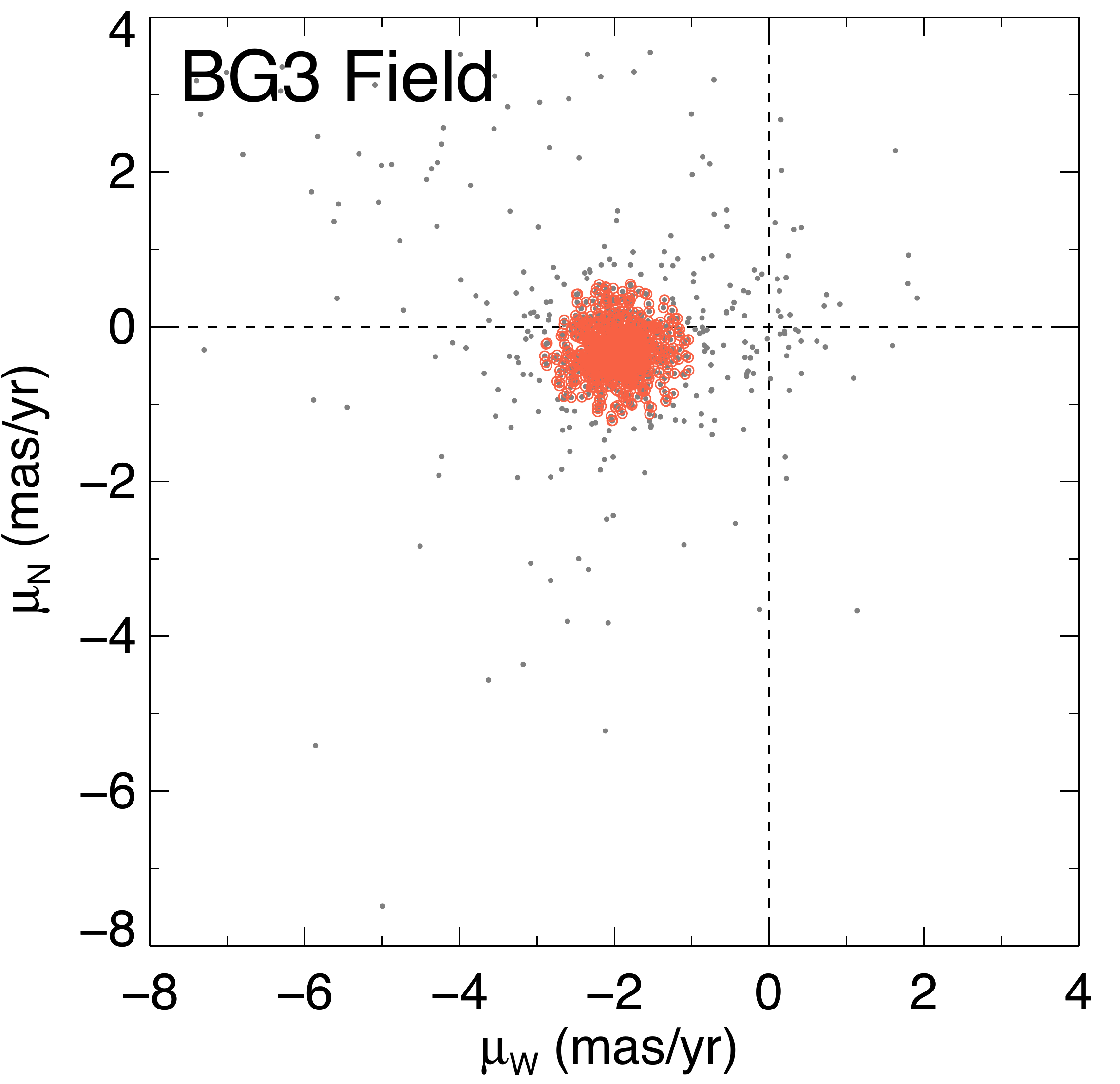}{0.33\textwidth}{}
            }

\caption{Color-magnitude (top) and proper motion diagrams (bottom) 
         for the three \hst\ fields. Stars selected as belonging to 
         the Magellanic Bridge are plotted in red while non-members 
         are plotted in gray. In the top panels, we overplot isochrones 
         with metallicities [Fe/H] $= -1.0$ and ages 0.25, 1, 3, 5, and 
         10 Gyr to represent stellar populations expected in these regions.
         Distances of 62, 62, and 50~kpc were adopted respectively for 
         BG1, BG2, and BG3. We applied reddening to the isochrones 
         based on the $E(B-V)$ values estimated from interpolating 
         the reddening maps of \citet{schlegel98}, and the total absorption 
         values were adopted from Table~6 of \citet{schlafly11}.
         \label{f:cmdpm}
         }
\end{figure*}
%

Figure~\ref{f:cmdpm} illustrates the selection of Magellanic Bridge 
stars in our \hst\ fields. The top panels show the CMDs, while the 
lower panels show the PM diagrams of all stars detected in the images. 
Selection of Magellanic Bridge stars in the target fields is 
straightforward since the PM diagrams exhibit conspicuous clumps as 
expected for groups of stars co-moving in the same direction. We first 
identified these clumps and selected candidate members of the 
Magellanic Bridge based on their distance from the average 
$(\muw, \mun)$ of the clumps. For this we define a local reference frame
for $\muw$ and $\mun$ with $\muw \equiv -(d\alpha / dt)\ cos(\delta)$ and 
$\mun \equiv d\delta / dt$. We then inspected the CMDs to verify 
that the majority of stars in the clump are consistent with an LMC- 
or SMC-like stellar population. The overlaid isochrones in 
the top panels of Figure~\ref{f:cmdpm} were adopted from the 
Dartmouth Stellar Evolution Database \citep[DSED,][]{dotter08a}, and 
represent such a population. Our goal here is not to carry out a 
detailed stellar population study for each field but to use the 
CMDs to select highly probable members of the Magellanic Bridge. 
With this in mind, we allowed a fairly wide range in color relative 
to the isochrones when selecting members, and only filtered out stars 
noticeably segregated in the CMD. Most of the non-members are far 
redder than the selected Magellanic Bridge candidates, and are most 
likely giant stars in the MW halo that happen to lie in the same 
region occupied by the Bridge stars in the PM diagram. 
We would add that all of the isochrone ages displayed represent populations formed before the most recent interaction between the SMC and LMC.
The average PMs of selected stars in each field were then calculated by 
taking the error-weighted mean, and the uncertainties of the 
averages were computed by propagating the individual PM uncertainties. 
We have also added the uncertainties originating from setting up 
the stationary reference frame using galaxy positions in quadrature, 
which typically dominates the final PM uncertainties. Results are 
shown in Table~\ref{t:pmresults}. We note that our results are 
insensitive to the CMD selection of Bridge stars. For example, we 
repeated our selection using a much more conservative criteria 
(i.e., only allowing stars consistent with the isochrones in 
Figure~\ref{f:cmdpm} within their color errors), and the resulting 
average PMs are all consistent with those in Table~\ref{t:pmresults} 
within their 1$\sigma$ uncertainty. We have also verified that there 
are no correlations between the locations in the CMDs and the 
PM diagrams for the selected Bridge stars. 

In addition to the three fields measured using the background galaxies, five additional fields were observed with the intent to use background quasars to measure the PMs \citep[e.g., as in][]{NK13}. The first epoch was observed in late 2014 as part of the original program and a new second epoch was observed in late 2017 as part of our \hst\ program GO-14775 (PI: van der Marel). However, the sample of spectroscopically-confirmed QSOs available at the time were very bright compared to the average Bridge star, and even though we designed our \hst\ observations with short and long exposures in order to try to mitigate this, due to the tension between avoiding saturating the bright quasar while still observing a sufficiently large number of stars in the fields, we were unable to successfully measure high-quality PMs for these five fields. The resultant errors were roughly on the order of 1 mas yr$^{-1}$, and are not competitive with the dataset compiled above.

%
\begin{deluxetable}{lccc}
\tablecaption{Proper Motion Average and Dispersion for the Magellanic
              Bridge Stars in the \hst\ Fields
              \label{t:pmresults}}
\tablehead{
 \colhead{}      & \colhead{$\muw$}     & \colhead{$\mun$}     & \colhead{}                             \\ 
 \colhead{Field} & \colhead{($\masyr$)} & \colhead{($\masyr$)} & \colhead{$N_{\star}$\tablenotemark{a}} 
}
\startdata
\hst-BG1         & $-1.638 \pm 0.052$ & $-1.421 \pm 0.052$ & 259 \\
\hst-BG2         & $-1.503 \pm 0.020$ & $-0.799 \pm 0.020$ & 177 \\
\hst-BG3         & $-1.960 \pm 0.013$ & $-0.326 \pm 0.013$ & 912 \\
\hline
\enddata
\tablenotetext{a}{Number of Magellanic Bridge stars included in the PM calculations.}
\end{deluxetable}
%

\section{Data Analysis \& Model Comparisons}\label{sec:analysis}

\subsection{Data Analysis}\label{ssec:data}

For our analysis, we need the motions of the stars relative to the Clouds, not just their absolute motions. However, as our sample stretches across tens of degrees on the night sky, simply subtracting the systemic motion of the SMC (chosen as the zero-point for the system) is incorrect as the projection of the systemic motion onto the plane of the sky will shift dramatically. To address this we correct for the viewing perspective at each star, as outlined in \cite{vdM02}, in addition to subtracting the systemic SMC motion ($\mu_{W}$ = $-$0.82 mas yr$^{-1}$ and $\mu_{N}$ = $-$1.21 mas yr$^{-1}$, \citealt{zivick18}; consistent with the PM found by \citealt{gaiahelmi18}). With all of the individual motions shifted into this standard frame, we then transform the positions and PM vectors into a Cartesian frame, as defined in \cite{gaiahelmi18}, to allow for consistent calculations of motion along the Bridge. We define the $x$-axis as the line connecting the kinematic centers of the SMC (($\alpha$, $\delta$) (J2000) = (16.25$^\circ$,$-$72.42$^\circ$)) and LMC (78.76$^\circ$, $-$69.19$^\circ$) with positive in the direction of the SMC. The arrangement of our sources in this reference frame can be seen in Figure \ref{fig:modelpos}. We use this reference frame in all later analyses and comparisons to models and refer to proper motions calculated in this way as "relative proper motions'' in the figures. This same process of viewing-perspective correction and transformation is applied to the PMs of the three \hst\ fields as well in addition to the systemic motion of the LMC at its kinematic center. 

\begin{figure}
\begin{center}
 \includegraphics[width=3.4in]{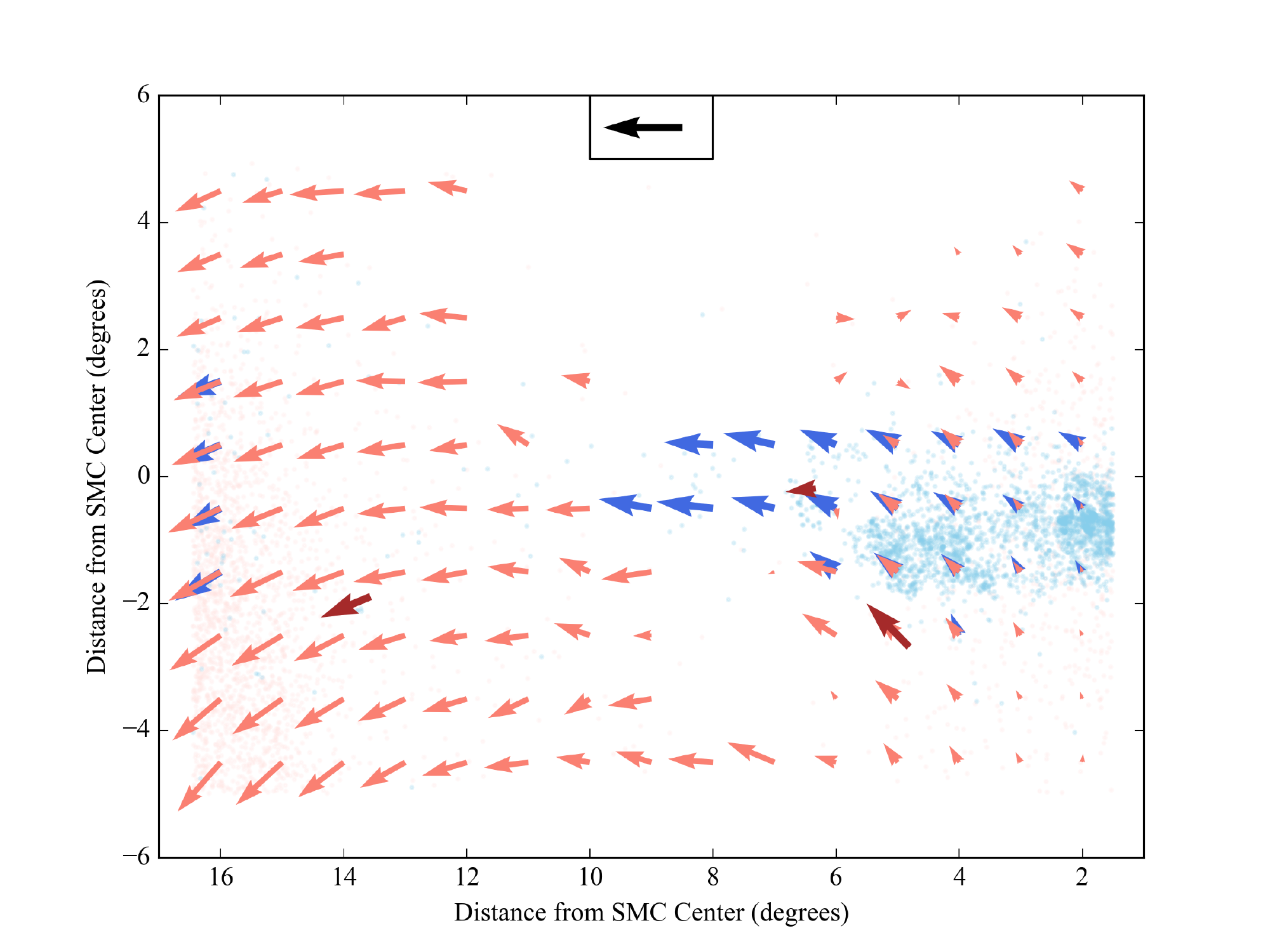} 
 \caption{Vector field of the residual PMs of the stellar populations in the Bridge relative to the SMC COM PM. The RG stars are displayed in orange-red, and the MS stars in blue. The \hst\ fields are marked in brown. The locations of each population are displayed in the background for reference. The median vectors are created from $0.5^{\circ} \times 0.5^{\circ}$ bins and are only calculated if five or more stars are present. A reference vector of 1 mas yr$^{-1}$ is provided at the top of the figure in black. The largest \gaia\ vector has a length of 1.01 mas yr$^{-1}$.}
   \label{fig:vecfield}
\end{center}
\end{figure}

In Figure \ref{fig:vecfield} we plot the resulting median residual PM vectors relative to the SMC center of mass (COM) PM, separated in $0.5^{\circ} \times 0.5^{\circ}$ degree bins across our selected region, with the two stellar populations indicated by our color convention. To help ensure that the displayed vectors are representative of the behavior at that location, only bins where there are five or more stars present are displayed. We see that the different stellar populations do not display significant differences in the vectors across the Bridge. However, we do see that when the absolute motion of the Small Magellanic Cloud is subtracted out, the residual Bridge motions display a general pattern of pointing away from the SMC towards the LMC. We display the measured motions for the \hst\ fields as well, which show a general agreement in the direction of motion, albeit different in the magnitude of the motion.

\begin{figure*}
\begin{center}
 \includegraphics[width=6.8in]{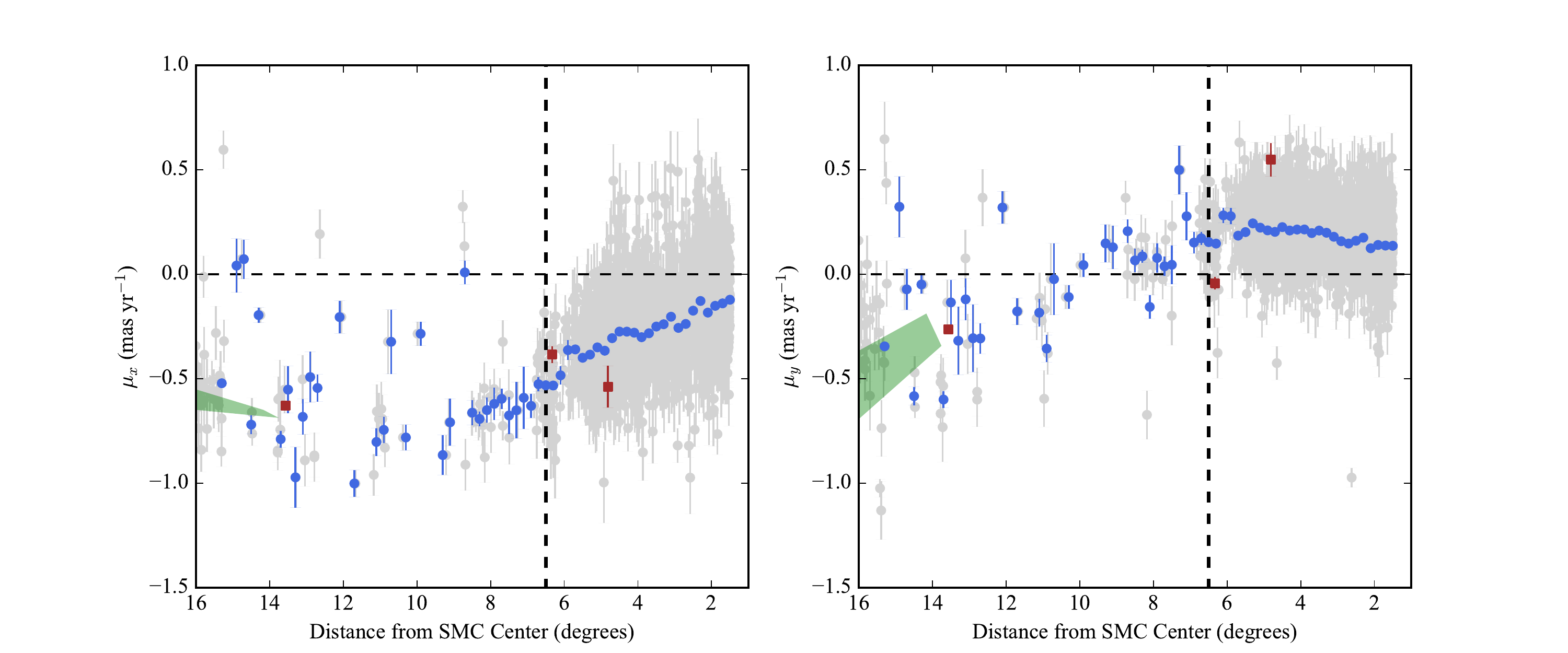} 
 \caption{(Left) Relative proper motions of the stars in the Bridge along the $x$-axis as a function of angular distance from the center of the SMC. All MS stars selected as part of the Bridge are displayed in gray. To understand the typical motion as a function of distance across the Bridge, the data are binned every 0.2 degrees, and the resulting error-weighted average PM in each bin is displayed in blue along with the standard error for weighted averages. The systematic errors of the \gaia\ DR2 catalog are not displayed. The motions of the \hst\ fields are marked in brown and the LMC-disk PMs by the light green region on the lefthand side of the plot. The vertical dashed line indicates the limit of comparison to Model 2, and the horizontal dashed line at 0 mas yr$^{-1}$ is a guide for the eye. (Right) Same as for the left plot but for the motion along the $y$-axis.}
   \label{fig:vtandat}
\end{center}
\end{figure*}

\begin{figure*}
\begin{center}
 \includegraphics[width=6.8in]{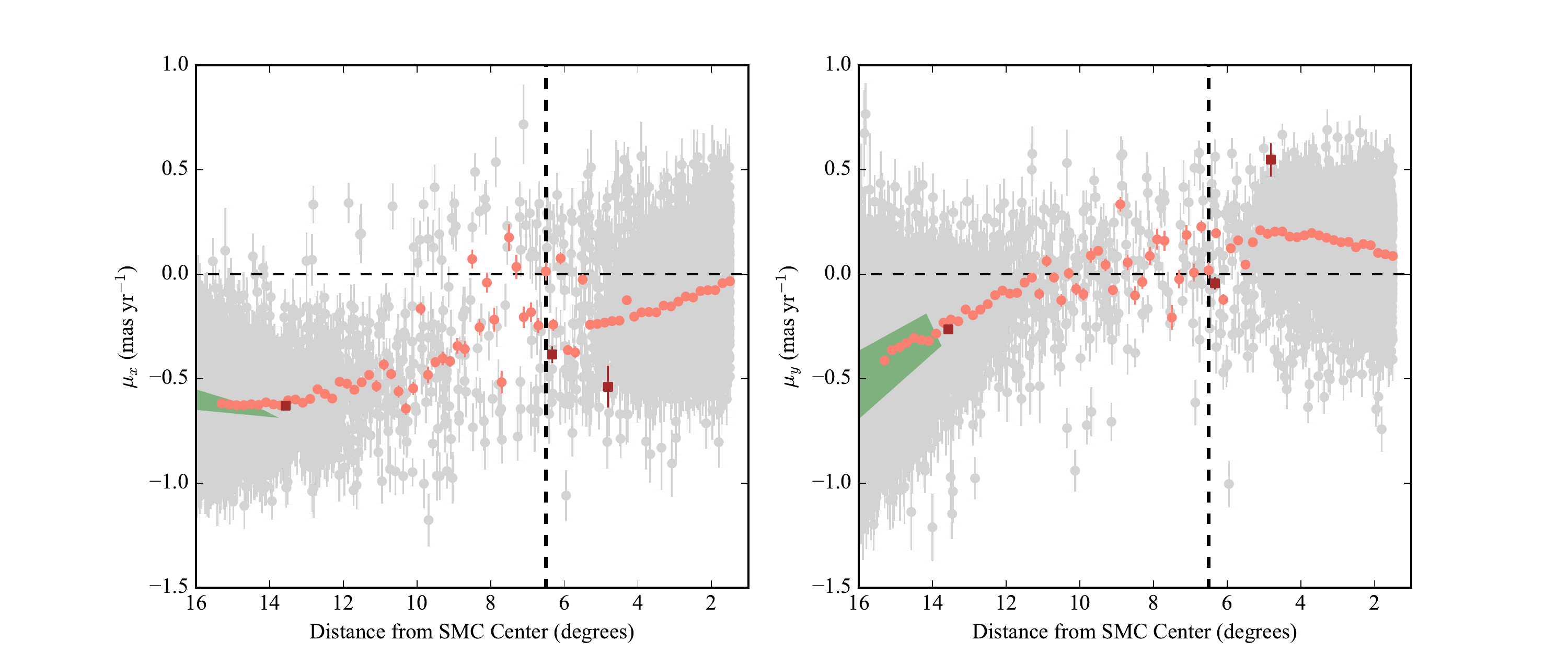} 
 \caption{(Left) Relative proper motions of the older stars in the Bridge along the $x$-axis as a function of angular distance from the center of the SMC. All RG stars selected as part of the Bridge are displayed in gray. To understand the typical motion as a function of distance across the Bridge, the data are binned every 0.2 degrees, and the resulting error-weighted average PM in each bin is displayed in orange-red along with the standard error for weighted averages. The systematic errors of the \gaia\ DR2 catalog are not displayed. The motions of the \hst\ fields are marked in brown and the LMC-disk PMs by the light green region on the lefthand side of the plot. The vertical dashed line indicates the limit of comparison to the models, and the horizontal dashed line is at 0 mas yr$^{-1}$ as a guide for the eye. (Right) Same as for the left plot but for the motion along the $y$-axis.}
   \label{fig:vtanrgb}
\end{center}
\end{figure*}

For the analysis, we keep all units in observed quantities, as converting to physical units, such as km s$^{-1}$, would require assumptions about the 3D structure of the Bridge. We found from our analysis that the \gaia\ parallaxes, while efficient at removing foreground stars, are not good enough to afford improved insights into the distances along the Bridge (median parallax errors of $ \sim 0.05$ mas for stars brighter than $G < 17$, where expected parallax at 50 kpc is $\sim 0.02$ mas). 
The resulting relative motions for the different stellar populations are shown in Figures \ref{fig:vtandat} and \ref{fig:vtanrgb}, and are discussed below.

Given the large number of stars in our samples, for display-purposes we group the data every 0.2 degrees. Within each group, we calculate the error-weighted average PM and the standard error of the weighted average. This error calculation only captures the random error of the measurements, not the spatially correlated systematic errors in the \gaia\ DR2 catalog, which \cite{lindegren18} finds to between $\sim 0.07$ mas yr$^{-1}$ for sources averaged over less than a degree and $\sim 0.03$ mas yr$^{-1}$ for sources averaged over $\sim10$ degrees or more.
These average PMs are marked in Figures \ref{fig:vtandat} and \ref{fig:vtanrgb} by the color points with the raw data plotted as the gray points in the background. We note that for each bin the errors are displayed but that for many of the bins the resulting standard error is smaller than the points. The `raw' data display roughly similar spreads in PM. Potential differences could readily be attributed to the difference in the spatial distribution of the two populations, with the MS stars relatively tightly clustered together while the RG stars are spread out over nearly ten degrees.

We additionally display a range of possible LMC-bound motions, drawn from the rotating disk model of the LMC from \cite{vdM14}, as a light green region.
The HST motions are shown as red squares in each Figure with their calculated errors, which illustrate the motion of older MS and turn-off stars.
Reassuringly, we see that for both the MS stars and RG stars the \hst\ motions agree quite well with the \gaia\ data. We note that the errors displayed are scaled the same for both \hst\ and \gaia\ so the comparable
precision of the \hst\ fields is real, despite the far fewer number of stars that have been averaged in each field. This illustrates that HST remains unique for small-field astrometric studies at faint magnitudes and large distances.

\subsection{H {\small I} and Stellar Comparisons}\label{ssec:hi}

\begin{figure}
\begin{center}
 \includegraphics[width=3.4in]{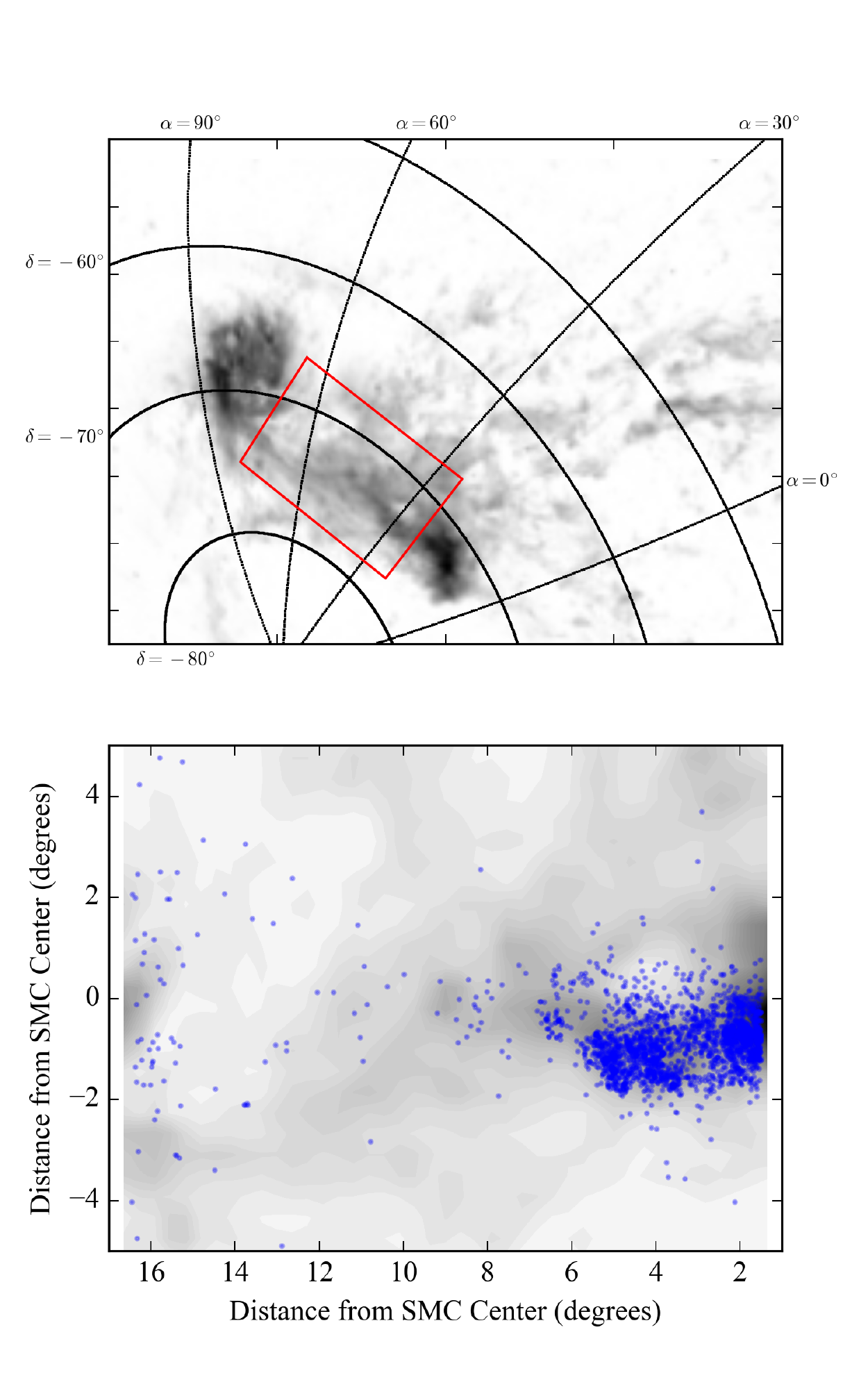}
 \caption{(Top) H {\small I} gas intensity map from \cite{putman03} with lines of constant RA and Dec provided for reference. The LMC is the large structure in the middle left of the panel and the SMC is located below and to the right of the LMC with the Bridge stretching between them. (Bottom) The H {\small I} map transformed into our working frame with the location of the selected MS stars overplotted in blue.}
   \label{fig:hi}
\end{center}
\end{figure}


As discussed briefly in Section \ref{ssec:gaiadata}, while two distinct MS branches are discernible in Figure \ref{fig:cmdsel}, the kinematic and spatial properties of the two branches are not significantly different. As such we choose to consider all MS stars together. 
For these young stars, we test for potential correlations with the H {\small I} gas distribution in the Bridge. For this comparison, we use the H {\small I} data from \cite{putman03}, and in Figure \ref{fig:hi} plot the gas intensity in addition to the locations of the MS stars. The correlation between the H {\small I} and the stars 
is immediately clear from the Figure, a trend that has been demonstrated in previous studies \citep[e.g.,]{skowron14}. We can see a large overlap of young stars with the dense arm of H {\small I} gas stretching out towards the LMC. We also note that slightly further out, at $\sim 8$ degrees, we observe a slight overdensity of young stars that falls between two peaks in the H {\small I} gas. Given the tight spatial correlation between the gas and the stars, we can infer that the behavior of these stars should indeed be similarly correlated with the kinematics of the underlying gas.

\begin{figure*}
\begin{center}
 \includegraphics[width=6.8in]{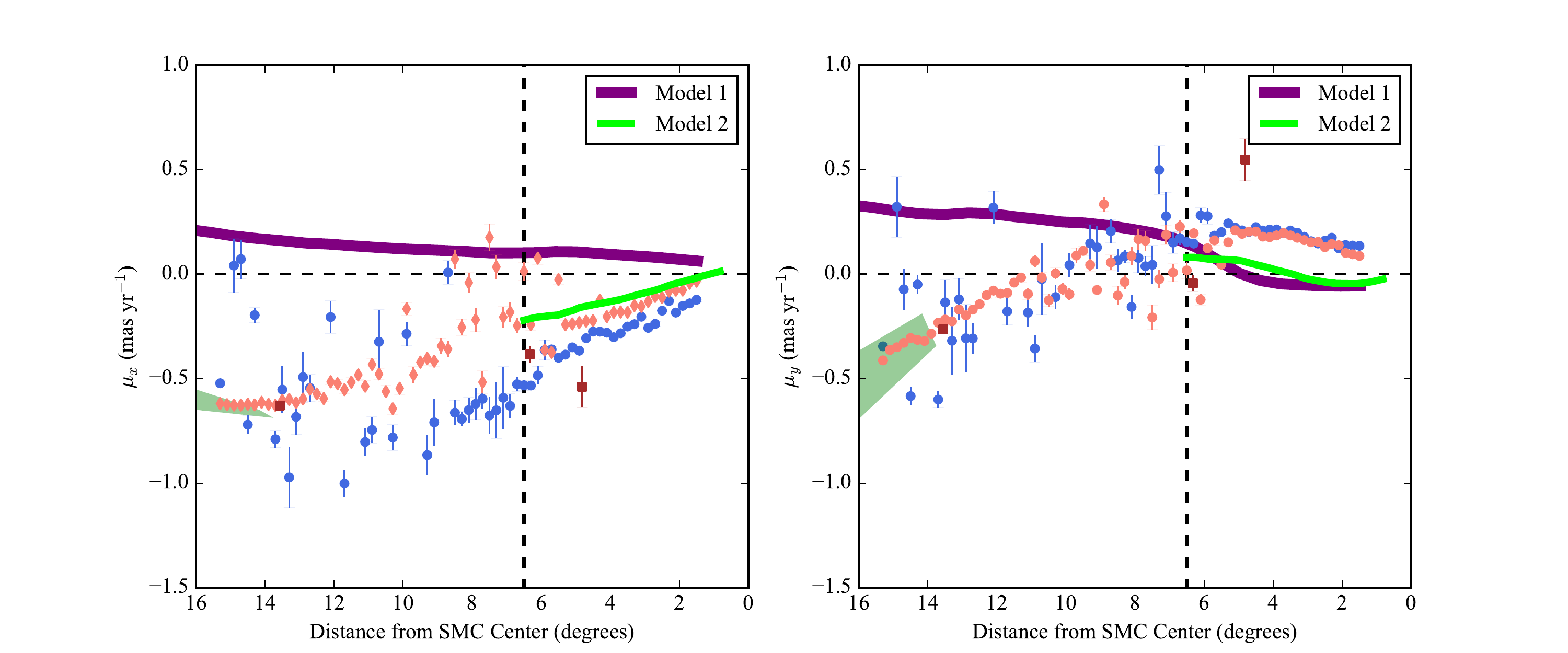} 
 \caption{(Left) Error-weighted average relative PMs along the Bridge in the $x$-axis direction, calculated as described in Section \ref{ssec:data} for the RG (orange-red) and MS (blue) stars. The motions of the \hst\ fields are marked by the brown squares and the LMC-disk PMs by the light green region on the lefthand side of the plot. The predicted motions from the two models are plotted here as well (purple for Model 1, lime-green for Model 2). The average error for the models is on the order of $\sim 0.1$ mas yr$^{-1}$. Model 2 allows for a direct collision between the SMC and LMC while Model 1 assumes they do not. At the start of the Bridge ($\sim2$ degrees from the SMC) Models 1 and 2 begin diverging, with Model 2 motions having a similar trend as both the observed RG and MS star motions. (Right) Same as the left plot but for motion along the $y$-axis. Near the start of the Bridge, Models 1 and 2 do not provide significant discriminating power. However, as Model 1 continues for the length of the Bridge, we observe a clear divergence from the data for both the RG and MS stars on approach to the LMC-side of the Bridge.}
   \label{fig:modelcomp}
\end{center}
\end{figure*}

Given the preferred age of tens of Myr for the MS stars and this tight correlation, we can interpret the two different populations as pre- and post-interaction with the LMC, as the RG stars are on the order of 1 Gyr old and the collision timeframe has been constrained to be roughly 100 Myr ago \citep{zivick18}. With this framework in mind, we look at the differences in behavior between the MS and RG stars, focusing on the weighted average PMs of each to compare the populations (shown against each other in Figure \ref{fig:modelcomp}).

In the $x$-direction, there appears to be a slight offset between the old and young populations with the MS stars having systematically larger negative PMs than the RG stars. Using the difference of the averages divided by the errors summed in quadrature as a statistic of significance, we find almost every bin before 7 degrees to be significant at the $3 \sigma$ level or greater. Even when accounting for the potential systematic error introduced by the spatial correlations (assumed to be $\sim 0.04$ mas yr$^{-1}$ given the intermediate spatial scales listed earlier), many of the individual bins still remain significant at the $3 \sigma$ level. Past 7 degrees the stellar sparsity makes statistical comparisons difficult, so we refrain from over-analyzing the trends. Interestingly in the $y$-direction we observe no such significance. Indeed across most of the Bridge, even in the sparse regions, the MS and RG populations appear to generally agree with each other. However, this is not an entirely unexpected result given the comparisons of the two Models, discussed further below. We do note the apparent structure in $\mu_{x}$ for the RG stars with a cluster of points above 0.0 mas yr$^{-1}$, stretching from $\approx$ 6 degrees to 10 degrees. However, further examination of these stars does not reveal any significant spatial correlations or correlations in $\mu_{y}$. One potential explanation would be that this is a detection of the RG tidal features of the SMC and LMC found in \cite{belokurov17}.

We also compare the location of the LMC-disk PMs to the data. In both the $x$- and $y$-directions we see the RG data matching well with the predicted PMs of the disk, though we note that this only holds true for near the LMC. Within $\sim 10$ degrees of the SMC, one observes a clear shift in the behavior of the stars. For the MS stars, the agreement is not as clear. The PMs in the $x$-direction appear to have a rough agreement, but there is a noticeable offset in the $y$-direction. We posit that the MS stars measured here originated from H {\small I} gas not initially belonging to the LMC as an explanation for this disparity, but given the sparsity of the data, refrain from attempting further analysis.


\subsection{Model Comparisons}\label{ssec:comp}

To understand the implications for the Magellanic system, we compare our data against simulations of the interactions between the Clouds from \cite{besla12}. Two models are explored, one in which the SMC and LMC 
interact tidally but remain relatively well-separated from each other ($\sim$ 20 kpc separation), referred to as Model 1, and one in which the SMC and LMC collide ($\sim$ 2 kpc separation), referred to as Model 2. In Model 1, the Bridge forms out of gas and stars tidally 
stripped from the SMC by the LMC. However, in Model 2 the SMC gas undergoes ram pressure stripping after encountering the LMC 
gas as it passes through the LMC's disk.  This hydrodynamic interaction enhances the density of the stripped gas and forces the corresponding stars that form in-situ to trace the SMC's  motion back towards the LMC.
From the presence of in situ star formation known already in the Bridge \citep[e.g.,][]{harris07}, we have reason to prefer the latter scenario, but our data allow us to further constrain the interaction history. For more details on the computational aspects of the simulations, please refer to \cite{besla12}.

The results from the simulations, similarly transformed and binned as our data, are displayed against the average PMs of the data in Figure \ref{fig:modelcomp} (Model 1 in purple, Model 2 in lime-green). We convert the physical units of the simulation (kpc, km s$^{-1}$) to observed quantities (degrees, mas yr$^{-1}$) to reduce the number of assumptions required for manipulating the data. For this conversion, we adjust the center of mass (COM) position of the modeled SMC to match the observed COM location of the SMC.
Note that the Bridge in Model 2 does not extend as far as in Model 1 (the area marked by the dashed black lines in Figure \ref{fig:modelpos} denote the area covered by Model 2, whereas Model 1 covers the entire area of the figure), limiting our ability to fully compare to our data.  
Nonetheless, the models do clearly predict distinct and different PM signals. Additionally, when we test limiting the spatial selection of our data for comparison to Model 2, we do not find any noticeable shifts in the average PMs for either the MS or RG populations. As a result, we choose to present kinematic information for all stars in the Bridge area. 
The two models diverge in the $x$-direction providing a clear test for comparison. The predicted motions in the $y$-direction are not as starkly different near the beginning of the Bridge, but we note that the continuation of Model 1 beyond $\sim$ 6 degrees from the SMC does provide some additional discriminatory power.

Before comparing the observed data to the simulated data, we note that the exact magnitudes of the motions are not a point of emphasis. Given the number of parameters involved in setting up the simulation, and with total LMC 
\& SMC masses being crucial unknowns in this, we do not expect that our data will perfectly replicate the predictions of the models. 
Instead we focus on comparisons of the trends in the data and the models to help provide a physical intuition for interpreting the data. That being said, perhaps surprisingly, we do find that the magnitudes of the PMs of the predicted and observed data along the Bridge do live in the same ballpark.

In comparing the data to Model 1 in Figure~\ref{fig:modelcomp}, we see a distinct disagreement between data and model in the $x$-direction. From the closest point in to the SMC, the values begin to diverge. In the $y$-direction, the difference is not as dramatic close to the SMC, but as the simulation data approaches the LMC, the predicted motion continues to increase in a positive direction while our observed data trends in the opposite direction, ending with a difference of almost 1 mas yr$^{-1}$. For Model 2, the predicted motions along the $x$-direction agree well with the observed data, although we are limited in the extent of our comparison beyond $\sim 6^{\circ}$ from the SMC center. However, this limitation itself provides a potential test as the shorter Bridge forms as a result of the direct collision and the resulting gas interactions between the SMC and LMC. Interestingly, we observe a distinct decline in the number of MS stars beginning around a similar distance into the Bridge as in Model 2. In the $y$-direction, we see a similar difference in the magnitudes of the motions as with Model 1, although not at as significant a level of disagreement, and the trend directions of both models and data roughly agree within $6^{\circ}$ of the SMC. 

In both models, the SMC is initially modeled as a rotating disk in a prograde orbit about the LMC, which enables the formation of the Magellanic Stream via tidal stripping. In Model 1, the lack of a direct collision means that the SMC disk retains ordered rotation. As a result, the tidally stripped material that forms the bridge contains residual signatures of the disk rotation, resulting in the positive motion along both the x and y direction in Figure 9.  In contrast, in Model 2, the SMC disk is destroyed in the collision (\citealt{besla11}; Besla et al. in prep). As such, both stripped stars and gas track the motion of the SMC back towards the LMC, without any rotation.
Given the known structure of the H {\small I} gas, and now the observed motions of stars moving away from the SMC, we find strong evidence for the scenario of a recent direct collision.



\section{Discussion \& Conclusions}\label{sec:dnc}

We present the first detailed analysis of the PM kinematics of the stellar component of the Magellanic Bridge using a combination of \gaia\ and \hst\ data. 
In the \gaia\ data we examine two different stellar populations, the MS and RG stars. In both cases, we use \gaia\ parallaxes, photometry, and kinematics to help discriminate between foreground stars and SMC/LMC stars. The \gaia-selected data span the entire length of the Bridge between the two Clouds. We point to the observable split between two main sequence populations to illustrate our ability to select a "clean'' sample of Magellanic stars.

Milky Way contamination is less of a concern with the \hst\ data. There we measure PMs in three Bridge fields, two relatively close to the SMC and one relatively close to the LMC. The PMs are measured with respect to background galaxies and over baselines of $\sim 4 - 13$ years. We pick up a much fainter, and relatively old population of MS and turn-off stars with \hst\ compared to \gaia\, as would be expected. One of the \hst\ field locations overlaps with the \gaia\ data, while the other two probe independent directions along the Bridge. The overlapping field gives us an opportunity for a direct comparison between \gaia\ PMs and \hst -measured PMs, albeit targeting different stellar populations, and these two independently-measured PM sets are found to be consistent with each other.

The different stellar populations probed by our datasets, in turn, give us an opportunity to investigate population-based structure and kinematics. The young MS stars display a strong spatial correlation with the underlying H {\small I} gas, unlike the RG stars that trace a broader dispersed structure around both the SMC and LMC. However, for the kinematics, both the RG and the MS stars exhibit similar behavior in increasing magnitude of their motion towards the LMC. The other component of their motion in the plane of the sky remains roughly consistent with the systemic motion of the SMC, only decreasing near the LMC. 

We compare the PM kinematics along the Bridge to predictions from two numerical simulations of the interaction-history of the Clouds from \cite{besla12}. The two different numerical simulations examined both consider the Bridge to be caused by tidal disturbance of the SMC by the LMC on a recent ($\sim100$ Myr) past encounter, but in Model 1, the Clouds remain relatively well-separated, with perhaps a grazing past encounter with an impact parameter of $\sim 20$ kpc, while in Model 2, the SMC goes directly through the LMC, with an impact parameter for the encounter of $\sim 2$ kpc (for reference, the LMC's disk radius is 18.5 kpc \cite{mackey16}). 
As such, Model 2 also allows for a hydrodynamic interaction between the SMC and LMC gas disks and ultimately destroys any signature of rotation in the SMC main body \citep{besla11}.
These two models predict different kinematic signatures in the $x$-direction, defined as the axis that lies along the line that connects the centers of the LMC and SMC (see Figure~\ref{fig:modelpos}), and when compared against the observational data, we find strong agreement with the direct collision model (Model 2). Combined with previous studies on the interaction parameters of the Clouds \citep[e.g.,][]{besla12,zivick18}, the growing body of evidence heavily favors such a direct collision \citep[e.g.,][]{oey18}, with an impact parameter of a few kpc.

Future work in this area will consist of continuing to draw in other types of data sets (e.g., star formation histories, metallicities)  to build a more holistic view of the history of the Clouds. This includes deeper examinations of the gas content of the Clouds where recent work has helped constrain the histories both using H {\small I} data \citep{mcclure18} and molecular gas \citep{fukui18}. Future data releases from \gaia\ will also continue to improve in data quality, but specifically, improvements in the parallaxes will allow us to include distances along the Bridge both as a constraint in the interaction history and more broadly to better separate out Magellanic debris (Bridge(s), Stream(s)) from Milky Way pollutants. Better distances for the Magellanic RGs will also aid a more rigorous investigation of population-based kinematic differences in the Bridge.

Additionally, analysis of the PM kinematics of the stellar populations of the SMC main body from \gaia\, along the lines of the analysis present here, will allow us to better constrain its geometry. At present, there is little evidence for internal rotation in the SMC, and strong evidence that the main body is being tidally disrupted, based largely on \hst\ data \citep{zivick18}. The addition of radial velocities will also add one more piece to the puzzle of the Magellanic Clouds, which are looking more and more like a local analog of the Antennae galaxies. As shown by Figure~\ref{fig:hi}, perhaps the most striking aspect of the data set presented here is the strong spatial correlation between H {\small I} gas in the Bridge and very young stars. Clearly the Clouds are an ideal laboratory to study star formation in a low metallicity regime.


On the numerical side, upcoming work will explore the impact of the LMC$-$SMC collision on the structure of the SMC main body (Besla et al. in prep).  Future studies including a more realistic treatment of star formation are needed to better understand the consequences of the recent violent interaction history to the star formation histories of the Clouds. 

Support for Program number GO-13834 was provided by NASA through a grant from the Space Telescope Science Institute, which is operated by the Association of Universities for Research in Astronomy, Incorporated, under NASA contract NAS5-26555.

This work has made use of data from the European Space Agency (ESA) mission
{\it Gaia} (\url{https://www.cosmos.esa.int/gaia}), processed by the {\it Gaia}
Data Processing and Analysis Consortium (DPAC,
\url{https://www.cosmos.esa.int/web/gaia/dpac/consortium}). Funding for the DPAC
has been provided by national institutions, in particular the institutions
participating in the {\it Gaia} Multilateral Agreement.

This project is part of the HSTPROMO
(High-resolution Space Telescope PROper MOtion)
Collaboration\footnote{http://www.stsci.edu/$\sim$marel/hstpromo.html},
a set of projects aimed at improving our dynamical understanding of
stars, clusters and galaxies in the nearby Universe through
measurement and interpretation of proper motions from HST, Gaia, and
other space observatories. We thank the collaboration members for the
sharing of their ideas and software.

\bibliography{Zivick_bib}{}

\end{document}